\documentclass[twocolumn]{jpsj2} 
\usepackage{bm}

\title{Quantum Dot Spin Filter in Resonant Tunneling and
Kondo Regimes}

\author{Mikio
\textsc{Eto}\thanks{E-mail address: eto@rk.phys.keio.ac.jp}
and Tomohiro \textsc{Yokoyama}}

\inst{Faculty of Science and Technology, Keio University,
3-14-1 Hiyoshi, Kohoku-ku, Yokohama 223-8522, Japan}

\abst{
A quantum dot with spin-orbit interaction can work
as an efficient spin filter if it is connected to $N$
($\ge 3$) external leads via tunnel barriers.
When an unpolarized current is injected to a quantum dot
from a lead, polarized currents are ejected to other
leads. A two-level quantum dot is examined as a minimal model.
First, we show that the spin polarization is
markedly enhanced by resonant tunneling when the
level spacing in the dot is smaller than the level
broadening. Next, we examine the many-body resonance induced
by the Kondo effect in the Coulomb blockade regime. A large spin
current is generated in the presence of the SU(4) Kondo effect
when the level spacing is less than the Kondo temperature.}

\kword{quantum dot, spin filter, spin-orbit interaction,
Kondo effect, spin Hall effect}

\begin{document}
\maketitle


The generation of spin current with no magnetic
field or ferromagnets is an important issue for
spin-based electronics, ``spintronics.''\cite{Zutic}
In this context, the spin-orbit (SO) interaction has
attracted much interest.
For conduction electrons in direct-gap semiconductors,
an external potential $U({\bm r})$ results in the
Rashba SO interaction\cite{Rashba,Rashba2}
\begin{equation}
H_\text{RSO} =
\frac{\lambda}{\hbar} {\bm \sigma} \cdot
\left[{\bm p} \times {\bm \nabla} U({\bm r}) \right],
\label{eq:SOorg}
\end{equation}
where ${\bm p}$ is the momentum operator and
${\bm \sigma}$ is the Pauli matrices indicating
the electron spin ${\bm s}={\bm \sigma}/2$.
The coupling constant $\lambda$ is markedly
enhanced by the band effect, particularly
in narrow-gap semiconductors, such as
InAs.\cite{Winkler,Nitta}
The spatial inversion symmetry is broken in
compound semiconductors, which gives rise to another
type of SO interaction, the Dresselhaus SO
interaction.\cite{Dresselhaus} It is given by
\begin{eqnarray}
H_\text{DSO}=\frac{\lambda'}{\hbar}
\bigl[ p_x(p_y^2-p_z^2)\sigma_x+
p_y(p_z^2-p_x^2)\sigma_y
\nonumber \\
+p_z(p_x^2-p_y^2)\sigma_z \bigr].
\label{eq:Dresselhaus}
\end{eqnarray}

In the presence of SO interaction,
the spin Hall effect (SHE) produces a spin current traverse
to an electric field applied by the bias voltage.
Two types of SHE have been
intensively studied. One is an intrinsic SHE,
which is induced by the drift motion of carriers in the
SO-split band structures.\cite{Murakami,Wunderlich,Sinova}
The other is an extrinsic SHE caused by the
spin-dependent scattering of electrons by
impurities.\cite{Dyakonov}
Kato {\it et al}.\ observed the spin accumulation at
sample edges traverse to the current,\cite{Kato}
which is ascribable to the extrinsic SHE with
$U({\bm r})$ being the screened Coulomb potential by
charged impurities in eq.\ (\ref{eq:SOorg}).\cite{Engel}
In our previous studies,\cite{EY1,YE1}
we theoretically examined the extrinsic SHE in semiconductor
heterostructures due to the scattering by an artificial
potential created by antidots, STM tips, and others.
The potential is electrically tunable. We showed that
the SHE is significantly enhanced by the resonant scattering
when the attractive potential is properly tuned.
We proposed a three-terminal spin-filter including a
single antidot.

In the present letter, we study the enhancement of the
``extrinsic SHE'' by resonant tunneling through a
quantum dot (QD) with a strong SO interaction, e.g.,
InAs QD.\cite{Igarashi,Fasth,Pfund,Takahashi}
The QD is connected to $N$ external leads via
tunnel barriers. In the QD, the number of
electrons can be tuned, one by one, owing to the
Coulomb blockade when the electrostatic potential is
changed by the gate voltage $V_\text{g}$. The current
through a QD shows a peak structure as a function of
$V_\text{g}$ (Coulomb oscillation).
We use the term SHE in the following meaning:
For $N \ge 3$, when an unpolarized current is injected
to the QD from a lead, polarized currents are ejected to
the other leads. In other words, the QD works as
a spin filter. First, we examine the SHE around the current
peaks, where the resonant tunneling takes place.
We show that the spin polarization is markedly
enhanced when the energy-level spacing
in the QD is smaller than the level broadening due to
the tunnel coupling to external leads.
Next, we examine the many-body resonance induced by the Kondo
effect in the Coulomb blockade regime with spin 1/2
in the QD. We obtain a large spin current in the presence of
the SU(4) Kondo effect when the level spacing is less than
the Kondo temperature.

\begin{figure}
\begin{center}
\includegraphics[width=4.5cm]{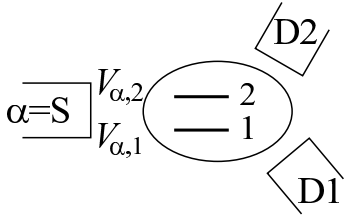}
\end{center}
\caption{Model for a quantum dot connected to
$N$ external leads ($N \ge 2$). The quantum dot has
two energy levels, $\varepsilon_j$ ($j=1,2$),
which are coupled to lead $\alpha$ by $V_{\alpha,j}$
[$\alpha=$S or D$n$, $n=1,\cdots, (N-1)$].
An unpolarized current is injected from lead S and
ejected to the other leads.
The spin-orbit interaction is present in the quantum dot.
}
\end{figure}

We assume that the SO interaction is present only in
the QD and that the level spacing in the QD is comparable
to the level broadening $\Gamma$ ($\sim 1$ $\mathrm{meV}$),
in accordance with experimental situations.\cite{Igarashi,
Fasth,Pfund,Takahashi}
The strength of SO interaction, $\Delta_\text{SO}$ in eq.\
(\ref{eq:Hdot}), is approximately
$0.2$ $\mathrm{meV}$.\cite{Fasth,Pfund,Takahashi}
As a minimal model, we examine two levels in the QD.
Note that previous theoretical
papers\cite{Kiselev3,Bardarson,Krich1,Krich2}
concerned the spin-current generation in a mesoscopic
region, or an open QD with no tunnel barriers,
in which many energy levels in the QD participate
in the transport.


We examine a two-level Anderson model with
$N$ $(\ge 2)$ leads, shown in Fig.\ 1.
The energy levels in the QD are denoted
by $\varepsilon_1$ and $\varepsilon_2$
before the SO interaction is turned on.
In the absence of magnetic field,
wavefunctions of the states, i.e.,
$\langle {\bm r} |1 \rangle$ and $\langle {\bm r} |2 \rangle$,
can be real.
In the case of Rashba SO interaction,
the orbital part in eq.\ (\ref{eq:SOorg}) is a pure imaginary
operator, and hence it has off-diagonal elements only;
$\langle 2| H_\text{RSO} | 1 \rangle =
\text{i} {\bm h}_\text{SO} \cdot {\bm \sigma}/2$
with $\text{i} {\bm h}_\text{SO}/2= (\lambda/\hbar) 
\langle 2 | ({\bm p} \times {\bm \nabla} U) | 1 \rangle$.
If the quantization axis of spin is taken in the
direction of ${\bm h}_\text{SO}$,
the Hamiltonian in the QD reads
\begin{eqnarray}
H_\text{dot}
&=& \sum_{\sigma=\pm 1}
(d_{1,\sigma}^{\dagger}, d_{2,\sigma}^{\dagger})
\left( \varepsilon_\text{d} - \frac{\Delta}{2}
\tau_z  + \sigma \frac{\Delta_\text{SO}}{2} \tau_y
\right)
\left( \hspace*{-0.2cm}
\begin{array}{c}
d_{1,\sigma} \\ d_{2,\sigma}
\end{array}
\hspace*{-0.2cm} \right)
\nonumber \\
& & +H_\text{int},
\label{eq:Hdot}
\end{eqnarray}
where $d_{j,\sigma}^{\dagger}$ and $d_{j,\sigma}$ are the
creation and annihilation operators of an electron with
orbital $j$ and spin $\sigma$, respectively.
$\varepsilon_\text{d} =(\varepsilon_1 + \varepsilon_2 )/2$,
$\Delta =\varepsilon_2 -\varepsilon_1$,
and $\Delta_\text{SO}=|{\bm h}_\text{SO}|$.
The Pauli matrices, $\tau_y$ and $\tau_z$, 
are introduced for the pseudo-spin representing level
$1$ or $2$. $H_\text{int}$
describes the Coulomb interaction between electrons.
The same form of the QD Hamiltonian is derived similarly
in the case of Dresselhaus SO interaction in eq.\
(\ref{eq:Dresselhaus}).\cite{com0}
Note that the level spacing
would be $\sqrt{\Delta^2+(\Delta_\text{SO})^2}$
in an isolated QD.

The state $|j \rangle$ in the QD is connected to lead
$\alpha$ by tunnel coupling, $V_{\alpha, j}$ ($j=1,2$),
which is real. The tunnel Hamiltonian is
\begin{eqnarray}
H_\text{T} & = &
\sum_{j=1,2} \sum_{\alpha,k,\sigma}
(V_{\alpha, j} d_{j,\sigma}^{\dagger}c_{\alpha k,\sigma}+
\text{h.c}.)
\nonumber \\
& = &
\sum_{\alpha,k,\sigma}
V_{\alpha} [ (e_{\alpha,1} d_{1,\sigma}^{\dagger}+
e_{\alpha,2} d_{2,\sigma}^{\dagger}) c_{\alpha k,\sigma}+
\text{h.c}.],
\label{eq:Htunnel}
\end{eqnarray}
where $c_{\alpha k,\sigma}$ annihilates an electron with
state $k$ and spin $\sigma$ in lead $\alpha$.
$V_{\alpha}=\sqrt{(V_{\alpha, 1})^2+(V_{\alpha, 2})^2}$
and $e_{\alpha,j}=V_{\alpha, j}/V_{\alpha}$. We
introduce a unit vector,
${\bm e}_{\alpha}=(e_{\alpha,1},e_{\alpha,2})^T$.
$V_{\alpha}$ is controllable by electrically tuning
the tunnel barrier, whereas ${\bm e}_{\alpha}$ is
determined by the wavefunctions
$\langle {\bm r} | 1 \rangle$ and
$\langle {\bm r} | 2 \rangle$ in the QD and hardly
controllable for a given current peak.
($\{ {\bm e}_{\alpha} \}$ and $\Delta$ vary
from peak to peak during the Coulomb oscillation.
We can choose a peak with appropriate parameters
for the SHE in experiments.)

We assume a single channel of conduction electrons
in the leads.
The total Hamiltonian is
\begin{equation}
H=\sum_{\alpha}\sum_{k,\sigma} \varepsilon_k
  c_{\alpha k,\sigma}^{\dagger} c_{\alpha k,\sigma}
  +H_\text{dot}+H_\text{T}.
\label{eq:Hamiltonian}
\end{equation}

The strength of the tunnel coupling is characterized by
the level broadening, $\Gamma_\alpha = \pi \nu_\alpha
(V_{\alpha})^2$,
where $\nu_\alpha$ is the density of states in lead $\alpha$.
We also introduce a matrix of
$\hat{\Gamma}=\sum_{\alpha} \hat{\Gamma}_{\alpha}$ with
\begin{eqnarray}
\hat{\Gamma}_{\alpha}
=
\Gamma_\alpha
\left( \begin{array}{cc}
(e_{\alpha, 1})^2 & e_{\alpha, 1}e_{\alpha, 2}
\\
e_{\alpha, 1}e_{\alpha, 2} & (e_{\alpha, 2})^2
\end{array}
\right).
\label{eq:broadening-M}
\end{eqnarray}

An unpolarized current is injected into the QD from a
source lead ($\alpha=$S) and output to other leads
[D$n$; $n=1,\cdots,(N-1)$]. The electrochemical potential
for electrons in lead S is
lower than that in the other leads by $-eV_\text{bias}$.
The current with spin $\sigma=\pm$ from lead $\alpha$ to
the QD is written as
\begin{equation}
I_{\alpha,\sigma}=\frac{\text{i}e}{\pi \hbar} \int d\varepsilon
\text{Tr}\{
\hat{\Gamma}_{\alpha}
[f_{\alpha}(\varepsilon)
(\hat{G}^\text{r}_{\sigma}-\hat{G}^\text{a}_{\sigma})+
\hat{G}^{<}_{\sigma}]
\},
\label{eq:current}
\end{equation}
where $\hat{G}^\text{r}_{\sigma}$,
$\hat{G}^\text{a}_{\sigma}$,
and $\hat{G}^{<}_{\sigma}$ are the retarded, advanced,
and lesser Green functions in the QD,
respectively, in $2\times 2$ matrix form
in the pseudo-spin space.\cite{Meir}
$f_{\alpha}(\varepsilon)$
is the Fermi distribution function in lead $\alpha$.
In the absence of electron-electron interaction,
$H_\text{int}$, the 
conductance into lead D$n$ with spin $\sigma$
is given by\cite{com1}
\begin{equation}
G_{n,\sigma}=
\left.
-\frac{\text{d}I_{\text{D}n,\sigma}}{\text{d}V_\text{bias}}
\right|_{V_\text{bias} = 0}
=\frac{4e^2}{h} \text{Tr}
[\hat{G}^\text{a}_{\sigma}(\varepsilon_\text{F})
\hat{\Gamma}_{\text{D}n}
\hat{G}^\text{r}_{\sigma}(\varepsilon_\text{F})
\hat{\Gamma}_{\text{S}}],
\label{eq:conductance0}
\end{equation}
where the QD Green function is
\begin{equation}
\hat{G}^\text{r}_{\pm}(\varepsilon) =
\left[
\left( \begin{array}{cc}
\varepsilon-\varepsilon_\text{d} + \frac{\Delta}{2}
 & \pm \text{i} \frac{\Delta_\text{SO}}{2}
\\
\mp \text{i} \frac{\Delta_\text{SO}}{2} &
\varepsilon-\varepsilon_\text{d} - \frac{\Delta}{2}
\end{array}
\right)
+\text{i} \hat{\Gamma}
\right]^{-1}
\label{eq:G0}
\end{equation}
and $\varepsilon_\text{F}$ is the Fermi energy.


\begin{figure}
\begin{center}
\includegraphics[width=7cm]{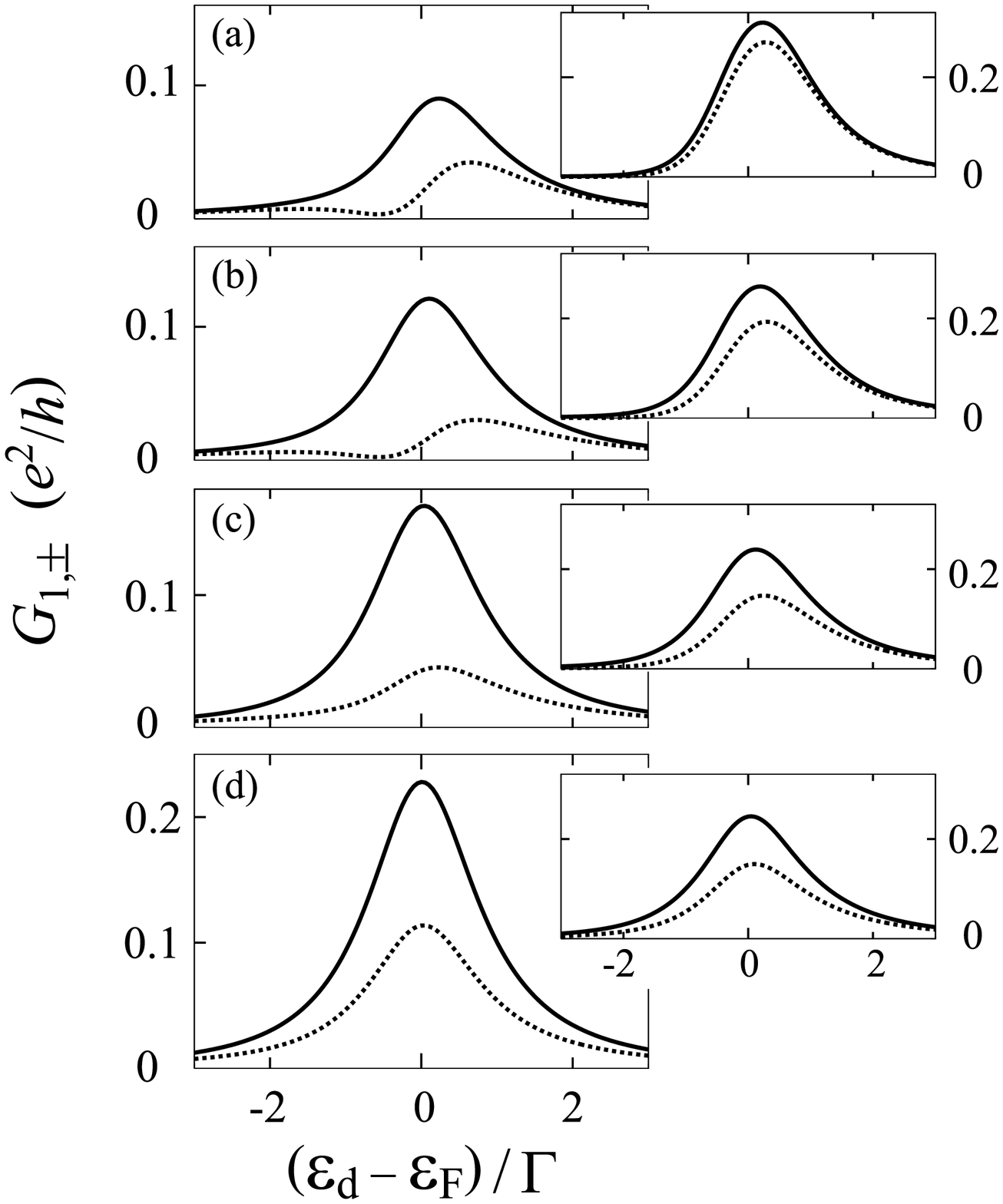}
\end{center}
\caption{
Calculated results of the conductance
$G_{1,\pm}$
as a function of energy level, $\varepsilon_\text{d}=
(\varepsilon_1 + \varepsilon_2 )/2$, in a
three-terminated quantum dot.
Solid (broken) lines indicate the conductance
$G_{1,+}$ ($G_{1,-}$) for spin
$\sigma=+1$ ($-1$).
The level broadening by the tunnel coupling to
leads S and D$1$ is
$\Gamma_\text{S} = \Gamma_{\text{D}1} \equiv \Gamma$
($e_{\text{S}, 1}/e_{\text{S}, 2} = 1/2$, 
$e_{\text{D}1, 1}/e_{\text{D}1, 2} = -3$),
whereas that to lead D$2$ is
(a) $\Gamma_{\text{D}2} =0.2\Gamma$,
(b) $0.5\Gamma$, (c) $\Gamma$, and (d) $2\Gamma$
($e_{\text{D}2, 1}/e_{\text{D}2, 2} = 1$).
$\Delta=\varepsilon_2 -\varepsilon_1=0.2\Gamma$
in the main panels and $\Delta=\Gamma$ in the insets.
$\Delta_\text{SO} =0.2\Gamma$.
}
\end{figure}

Now, we discuss the SHE in the vicinity of the
Coulomb peaks. The electron-electron interaction is
neglected in this regime.
From eqs.\ (\ref{eq:conductance0}) and
(\ref{eq:G0}), we obtain
\begin{eqnarray}
G_{n,\sigma} & = & \frac{e^2}{h}
\frac{4\Gamma_\text{S} \Gamma_{\text{D}n}}{|D|^2}
\left[ g_{n}^{(1)} + g_{n,\sigma}^{(2)} \right],
\label{eq:conductance}
\\
g_{n}^{(1)} & = &
\Biggl[
\left(
\varepsilon_\text{F}-\varepsilon_\text{d}-\frac{\Delta}{2}
\right)
e_{\text{D}n,1} e_{\text{S},1}
\nonumber \\
& &
+
\left(
\varepsilon_\text{F}-\varepsilon_\text{d}+\frac{\Delta}{2}
\right)
e_{\text{D}n,2} e_{\text{S},2}
\Biggr]^2,
\\
g_{n,\pm}^{(2)} & = &
\Biggl[
\pm \frac{\Delta_\text{SO}}{2}
({\bm e}_\text{S} \times {\bm e}_{\text{D}n})_z
\nonumber \\
& &
+
\sum_{\alpha} \Gamma_{\alpha}
({\bm e}_{\text{D}n} \times {\bm e}_{\alpha})_z
({\bm e}_\text{S} \times {\bm e}_{\alpha})_z
\Biggr]^2,
\end{eqnarray}
where $D$ is the determinant of
$[\hat{G}^\text{r}_{\sigma}(\varepsilon_\text{F})]^{-1}$
in eq.\ (\ref{eq:G0}), which is independent of $\sigma$,
and $({\bm a} \times {\bm b})_z = a_1 b_2 - a_2 b_1$.
Let us consider two simple cases. (I)
When
$\Delta \gg \Gamma_{\alpha}$ and $\Delta_\text{SO}$, 
$G_{n,\sigma}$ consists of
two Lorentzian peaks as a function of
$\varepsilon_\text{d}$, reflecting the resonant
tunneling through one of the energy levels,
$\varepsilon_{1,2}=
\varepsilon_\text{d} \mp \Delta/2$:
\begin{equation}
G_{n,\sigma} \approx \frac{4 e^2}{h}
\Gamma_\text{S} \Gamma_{\text{D}n}
\sum_{j=1,2}
\frac{(e_{\text{D}n,j} e_{\text{S},j})^2}
{(\varepsilon_j-\varepsilon_\text{F})^2+(\Gamma_{jj})^2},
\label{eq:G-largeD}
\end{equation}
where $\Gamma_{jj}$ [$jj$ component of
matrix $\hat{\Gamma}$;
$\sum_{\alpha} \pi \nu_{\alpha} (V_{\alpha,j})^2$]
is the broadening of level $j$. In this case,
the spin current [$\propto (G_{n,+}-G_{n,-})$] is very
small. $\Delta$ should be comparable to
or smaller than the level broadening
to observe a considerable spin current.
(II) In a two-terminated
QD ($N=2$), the second term in $g_{n,\pm}^{(2)}$
vanishes.
Since $g_{n,+}^{(2)}=g_{n,-}^{(2)}$,
no spin current is generated.\cite{Eto05b}
Three or more leads are required to observe a spin current,
as pointed out by other groups.\cite{Krich1,Zhai,Kiselev3}

We focus on $G_{1,\pm}$ in the three-terminal
system ($N=3$). Then
$g_{1,\pm}^{(2)} =
[\pm (\Delta_\text{SO}/2)
({\bm e}_\text{S} \times {\bm e}_{\text{D}1})_z +
\Gamma_{\text{D}2}
({\bm e}_{\text{D}1} \times {\bm e}_{\text{D}2})_z
({\bm e}_\text{S} \times {\bm e}_{\text{D}2})_z
]^2$.
We exclude specific situations in which two out of
${\bm e}_\text{S}$,
${\bm e}_{\text{D}1}$, and ${\bm e}_{\text{D}2}$ are
parallel to each other hereafter.
The conditions for a large spin current are as follows:
(i) $\Delta^{<}_{\sim}$ (level broadening),
as mentioned above. Two levels in the QD should
participate in the transport.
(ii) The Fermi level in the leads is close to the energy
levels in the QD,
$\varepsilon_\text{F} \approx \varepsilon_\text{d}$
(resonant condition).
(iii) The level broadening by the tunnel coupling to
lead D$2$, $\Gamma_{\text{D}2}$, is comparable to
the strength of SO interaction $\Delta_\text{SO}$.

Figures 2 and 3 show two typical results of
the conductance $G_{1,\pm}$ as a function of
$\varepsilon_\text{d}$ (Coulomb peak).
In $g_{1}^{(1)}$, $e_{\text{D}1,1} e_{\text{S},1}$ and
$e_{\text{D}1,2} e_{\text{S},2}$ have different
(same) signs in Fig.\ 2 (Fig.\ 3):
$g_{1}^{(1)}=0$ has no solution (a solution) in
$-\Delta/2 <
\varepsilon_\text{d}-\varepsilon_\text{F}
< \Delta/2$.

In Fig.\ 2, the conductance shows a single peak. 
We set
$\Gamma_\text{S}=\Gamma_{\text{D}1} \equiv \Gamma$.
When $\Delta=0.2 \Gamma$ (main panels),
we obtain a large spin current around the current
peak, which clearly indicates an enhancement of the SHE
by resonant tunneling [conditions (i) and (ii)].
With increasing $\Gamma_{\text{D}2}$
from (a) $0.2\Gamma$ to (d) $2\Gamma$, the spin
current increases first, takes a maximum in panel (c),
and then decreases [condition (iii)].
Therefore, the SHE is tunable by changing the tunnel coupling.
When $\Delta=\Gamma$ (insets), the SHE is
less effective, but we still observe a spin polarization of
$P=(G_{1,+}-G_{1,-})/(G_{1,+}+G_{1,-}) \approx 0.25$
at the conductance peak in panel (c).

\begin{figure}
\begin{center}
\includegraphics[width=5cm]{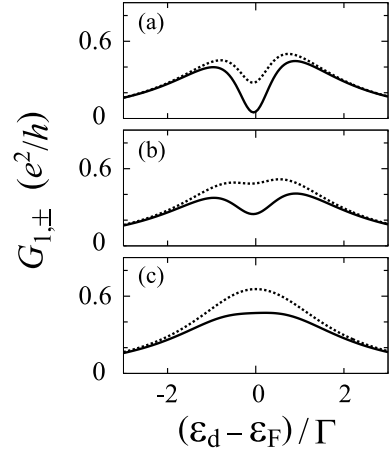}
\end{center}
\caption{
Calculated results of the conductance
$G_{1,\pm}$
as a function of energy level, $\varepsilon_\text{d}=
(\varepsilon_1 + \varepsilon_2 )/2$, in a
three-terminated quantum dot.
Solid (broken) lines indicate the conductance
$G_{1,+}$ ($G_{1,-}$) for spin
$\sigma=+1$ ($-1$).
The level broadening by the tunnel coupling to
leads S and D$1$ is
$\Gamma_\text{S} = \Gamma_{\text{D}1} \equiv \Gamma$
($e_{\text{S}, 1}/e_{\text{S}, 2} = 1/2$, 
$e_{\text{D}1, 1}/e_{\text{D}1, 2} = 3$),
whereas that to lead D$2$ is
(a) $\Gamma_{\text{D}2} =0.2\Gamma$,
(b) $0.5\Gamma$, and (c) $\Gamma$
($e_{\text{D}2, 1}/e_{\text{D}2, 2} = -1$).
$\Delta=\varepsilon_2 -\varepsilon_1=0.5\Gamma$.
$\Delta_\text{SO} =0.2\Gamma$.
}
\end{figure}

In Fig.\ 3, the conductance $G_{1,\pm}$ shows a dip
at $\varepsilon_\text{d} \approx \varepsilon_\text{F}$
for small $\Gamma_{\text{D}2}$.\cite{com3}
Around the dip,
the spin polarization is markedly enhanced:
$|P|$ is close to unity in panel (a).


Next, we examine the Kondo effect in the Coulomb blockade
regime with a single electron in the QD.
The Kondo effect is not broken by the SO interaction since
the time-reversal symmetry holds. For the electron-electron
interaction in the QD, we assume that
$H_\text{int}=U \sum_{j} n_{j,+} n_{j,-}
+U' \sum_{\sigma,\sigma'} n_{1,\sigma} n_{2,\sigma'}$,
where $n_{j,\sigma}=d_{j,\sigma}^{\dagger}d_{j,\sigma}$,
with infinitely large $U$ and $U'$.
The Kondo effect creates the many-body
resonant state at the Fermi level, and thus condition
(ii) is always satisfied. The resonant width is
given by the Kondo temperature $T_\text{K}$.\cite{Hewson}
When $T_\text{K} < \Delta$, the upper level in the QD
is irrelevant. The spin at the lower level is
screened out by the conventional SU(2) Kondo effect.
When $T_\text{K} > \Delta$, on the other hand, the
pseudo-spin as well as the spin are screened by
the SU(4) Kondo effect.\cite{com4}
The latter situation is required for an enhanced SHE
since two levels should be relevant to the resonance
[condition (i)].

The crossover between the SU(2) and SU(4) Kondo effects
can be semiquantitatively described by the slave-boson
mean-field theory.\cite{Lim}
The theory describes the Kondo resonant state on the
assumption of its presence and Fermi liquid behavior
and yields the conductance at temperature $T=0$.
A boson operator $b$ is
introduced to represent an empty state in the
QD. $d_{j,\sigma}^{\dagger}=
f_{j,\sigma}^{\dagger}b$ and $d_{j,\sigma}=
b^{\dagger} f_{j,\sigma}$, with a fermion
operator $f_{j,\sigma}$ representing the pseudo-spin $j$
and spin $\sigma$.
$H_\text{int}$ is taken into account by the constraint of
$Q \equiv \sum_{j,\sigma} f_{j,\sigma}^{\dagger}f_{j,\sigma}+
b^{\dagger}b-1=0$. $b$ is replaced with the mean field
$\langle b \rangle$,
which is determined by minimizing $\langle H+\lambda Q \rangle$
with the Lagrange multiplier $\lambda$.\cite{Hewson}
The conductance is given by eq.\ (\ref{eq:conductance})
if $\varepsilon_\text{d}$ and $\Gamma_\alpha$ are
replaced with the renormalized ones,
$\varepsilon_\text{d}+\lambda$
($\sim \varepsilon_\text{F}$) and
$\Gamma_\alpha \langle b \rangle^2$
($\sim T_\text{K}$), respectively.

Figure 4 shows $G_{1,\pm}$
as a function of $\varepsilon_\text{d}$ in the
three-terminal system.
The parameters are the same as those in the main panels
in Fig.\ 2. In the two-terminal situation (curve $a$;
$G_{1,+}=G_{1,-}$), the conductance increases
with decreasing $\varepsilon_\text{d}$
and saturates, indicating the charge fluctuation
regime and Kondo regime, respectively.
With three leads (curves $b$--$e$), we observe a
spin current around the beginning of the
Kondo regime. $P \approx 0.5$ in the case of
curve $e$.
As $\varepsilon_\text{d}$ is decreased further,
$T_\text{K}$ decreases and becomes smaller
than $\Delta$, which weakens the SHE.
We obtain similar results using the parameters
in Fig.\ 3.

\begin{figure}
\begin{center}
\includegraphics[width=7cm]{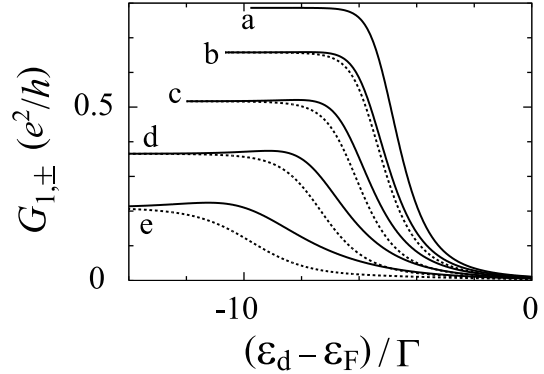}
\end{center}
\caption{
Calculated results of the conductance
$G_{1,\pm}$
as a function of energy level, $\varepsilon_\text{d}=
(\varepsilon_1 + \varepsilon_2)/2$, in a
three-terminated quantum dot in the presence of Kondo effect.
Solid (broken) lines indicate the conductance
$G_{1,+}$ ($G_{1,-}$) for spin
$\sigma=+1$ ($-1$).
The level broadening by the tunnel coupling to
lead D$2$ is
$\Gamma_{\text{D}2} =0$ (curve $a$; solid and
broken lines are overlapped),
$0.2\Gamma$ ($b$), $0.5\Gamma$ ($c$),
$\Gamma$ ($d$), and $2\Gamma$ ($e$).
The other parameters are the same as those in the main
panels in Fig.\ 2.
}
\end{figure}


In summary, we have examined the SHE in a
multiterminated QD with SO interaction.
The spin polarization in the output currents is
markedly enhanced by resonant tunneling if
the level spacing in the QD is smaller than the level
broadening. The spin current is also enlarged by
the many-body resonance due to the SU(4) Kondo effect.
The SHE is electrically tunable by changing the tunnel
coupling to the leads.

Hamaya {\it et al}.\ fabricated InAs QDs connected
to ferromagnets.\cite{Hamaya} If a ferromagnet is used
as a source lead in our model, spin filtering is
electrically detected through an ``inverse SHE.''
The current to lead D$1$ is proportional to
$(1+p\cos\theta) G_{\text{D}1,+}+
(1-p\cos\theta) G_{\text{D}1,-}$, where $p$ is
the polarization in the ferromagnet and $\theta$ is the
angle between the magnetization and ${\bm h}_\text{SO}$.

The SHE in QDs is useful for the fundamental research
as well as for the application to an efficient spin filter.
The SHE enhanced by resonant scattering or Kondo resonance
was examined for metallic systems with magnetic
impurities.\cite{Fert,Fert2,Guo}
In semiconductor QDs, we can observe the SHE due to the
scattering by a single ``impurity'' with the tuning of
various conditions.

The authors acknowledge fruitful discussion with
G.\ Sch\"on and Y.\ Utsumi.
This work was partly supported by a Grant-in-Aid for Scientific
Research from the Japan Society for the Promotion of Science,
and by the Global COE Program ``High-Level Global Cooperation for
Leading-Edge Platform on Access Space (C12).''
T.\ Y.\ is a Research Fellow of the Japan Society for
the Promotion of Science.


\begin{thebibliography}{99}
\bibitem{com3}
The conductance dip is caused by the destructive
interference between propagating waves through
two orbitals in the QD. 
In the two-terminal system without SO interaction,
the conductance $G_{1,\sigma} \propto g_{1}^{(1)}$
completely vanishes at the dip, where the
``phase lapse'' of the transmission phase takes
place. See ref.\ 36 and related
references cited therein.
\bibitem{com4}
We have two sets of Kramers' degenerate levels in the
QD and $N$ channels in the leads. By the unitary
transformation for the $N$ channels, we obtain
two channels connected to the QD and $(N-2)$
channels disconnected from the QD.
This is the situation of the full-screening Kondo
effect.\cite{Nozieres}
When $\Delta < T_\text{K}$, the SU(4) Kondo
effect is realized even when the strengths of the tunnel
coupling are not identical between the two levels in the QD.
This is because the fixed point is marginal (and
thus not unstable) in the Kondo scaling.\cite{Eto05}
\end{thebibliography}
\end{document}